\documentclass[pra,reprint,twocolumn,superscriptaddress,floatfix,aps,10pt,nofootinbib]{revtex4-1}

\usepackage{amssymb,amsmath,latexsym,bm,dsfont,graphicx,xcolor}
\usepackage{newtxtext}
\makeatletter\def\@pacs@name{DOI: }\makeatother

\renewcommand{\vec}[1]{{\bm{\mathrm{#1}}}}
\newcommand{\vhat}[1]{\hat{\bm{\mathrm{#1}}}}
\newcommand{\uda}{{\uparrow\downarrow}}
\let\epsilon\varepsilon
\let\Re\undefined
\let\Im\undefined
\DeclareMathOperator{\Im}{Im}
\DeclareMathOperator{\Re}{Re}
\DeclareMathOperator{\Tr}{Tr}

\DeclareMathOperator{\csch}{csch}

\begin{document}
\title{Spin transparency for an interface of an ultrathin magnet within the spin dephasing length}
\author{Kyoung-Whan Kim}%
\email{kwk@kist.re.kr}
\affiliation{Center for Spintronics, Korea Institute of Science and Technology, Seoul 02792, Korea}%
\date{\today}

\begin{abstract}
We examine a modified drift-diffusion formalism to describe spin transport near an ultrathin magnet whose thickness is similar to or less than the spin dephasing length. Most of the previous theories on spin torque assume the transverse component of a injected spin current dephases perfectly thus are fully absorbed into the ferromagnet. However, in the state-of-art multilayer systems under consideration of recent studies, the thicknesses of ferromagnets are on the order of or less than a nanometer, thus one cannot safely assume the spin dephasing to be perfect. To describe the effects of a finite dephasing rate, we adopt the concept of transmitted mixing conductance, whose application to the drift-diffusion formalism has been limited. For a concise description of physical consequences, we introduce an effective spin transparency. Interestingly, for an ultrathin magnet with a finite dephasing rate, the spin transparency can be even enhanced and there arises a non-negligible field-like spin-orbit torque even in the absence of the imaginary part of the spin mixing conductance. The effective spin transparency provides a simple extension of the drift-diffusion formalism, which is accessible to experimentalists analyzing their results.
\end{abstract}

\pacs{\quad}

\maketitle

\section{Introduction\label{Sec:introduction}}

Spin torque~\cite{Slonczewski1996,Berger1996,Ralph2008} has been a central concept in magnetism for a few decades, as it allows electrical control of magnetism. When a spin current is injected to a ferromagnet, its transverse component to magnetization dephases rapidly, thus its angular momentum is transferred to the magnetization, giving rise to a torque~\cite{Waintal2000,Heide2001,Zhang2002,Stiles2002}. It is typically assumed that the spin dephasing in the ferromagnetic bulk is infinitely fast, thus the spin current right at the interface solely determines the total angular momentum transfer to the ferromagnet~\cite{Shpiro2003}. Indeed, the spin dephasing length is on the order of or less than a nanometer~\cite{Zhang2004,Ghosh2012,Balaz2016}, this assumption has provided a very simple but still reasonable way to calculate a spin torque.

\begin{figure}
\includegraphics[width=7.0cm]{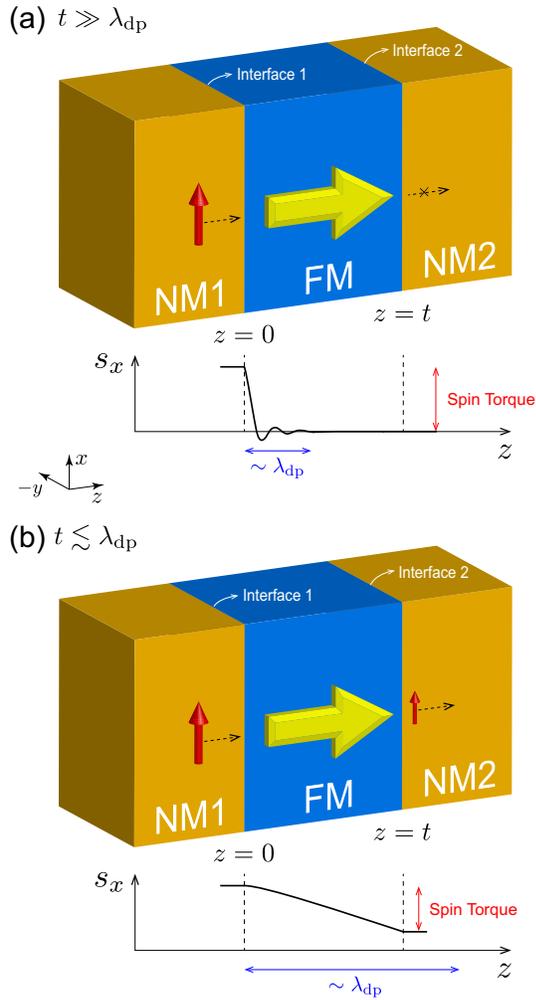}
\caption{Transport of transverse spin current through ferromagnets (FMs) (a) thicker than the spin dephasing length ($\lambda_{\rm dp}$) and (b) thinner than the spin dephasing length. For the former case, the two interfaces hardly communicate with each other, thus the total spin torque to the magnetic layer is determined solely by the spin current at $z=0^-$. In contrast, for the latter case, the spin chemical potential at $z=0^-$ can generate a spin current at $z=t^+$, thus the spin torque to the magnetic layer is no longer solely determined by the spin current at $z=0^-$. In this illustration, we denote the magnetic layer by a ferromagnet, but the argument is valid generally for antiferromagnets and ferrimagnets, for instance.
\label{Fig:diff}}
\end{figure}

Theoretically, the spin current at the interface is usually obtained by the drift-diffusion formalism~\cite{Valet1993,Zhang2002,Shpiro2003} with imposing proper boundary conditions (BCs). Considering a ferromagnet much thicker than the spin dephasing length, as illustrated in Fig.~\ref{Fig:diff}(a), one may assume that the transverse component of an injected spin current at the interface 1 does not reach the interface 2. Hence, as far as transverse spin transport is concerned, the two interfaces do not communicate with each other. Therefore, in the normal metal side near an interface, the transverse spin current at the interface, say $z=z_0$, is solely determined by the nonequilibrium spin chemical potential at the interface. Their relation is given by the celebrated magnetoelectronic circuit theory~\cite{Brataas2000,Brataas2001,Brataas2006}:
\begin{align}\label{Eq:conventional BC0}
e\vec{j}_s(z_0)&=\Re[G^\uda]\vec{m}\times[\vec{m}\times\vec{\mu}_s(z_0)]\nonumber\\
&\quad+\Im[G^\uda]\vec{m}\times\vec{\mu}_s(z_0).
\end{align}
Here $G^\uda$ is the spin mixing conductance of the interface, $e>0$ is the electron charge, $\vec{j}_s$ is the transverse spin current flowing to the normal metal side, $\vec{\mu}_s$ is the nonequilibrium spin chemical potential in the normal metal side, and $\vec{m}$ is the unit vector along magnetization in the magnetic layer. $\Re[G^\uda]$ and $\Im[G^\uda]$ are the coefficients for the Slonczewski-like torque [$\vec{m}\times(\vec{m}\times\vec{\mu}_s)$]~\cite{Slonczewski1996} and the field-like torque ($\vec{m}\times\vec{\mu}_s$)~\cite{Xiao2008,Oh2009}, respectively. With the BC [Eq.~(\ref{Eq:conventional BC0})], the spin drift-diffusion equation determines the spatial profiles of $\vec{\mu}_s(z)$ and $\vec{j}_s(z)$ self-consistently, and the spin torque to the ferromagnet is then calculated by the spin current right at the interface. The drift-diffusion formalism with the BC in Eq.~(\ref{Eq:conventional BC0}) has been used for numerous theories~\cite{Chen2013,Haney2013,Taniguchi2015,Amin2016,Amin2016a,Lee2013,Zhang2016} and experiments~\cite{Nakayama2012,Feng2012,RojasSanchez2014,Zhang2015,Nguyen2016,Qiu2016}.

A relatively intuitive way to understand the self-consistent procedure is introducing the spin transparency~\cite{Zhang2015} of a given interface.\footnote{Here, we denoted the reported `transparency' by the `\emph{spin} transparency', to emphasize that it is irrelevant for \emph{charge} transport. For instance, the spin transparency is not necessarily zero for a magnetic insulator, through which a charge current cannot flow. Still, it is worth noting that the spin transparency considered throughout this paper is the one for the transverse spin transport, not the longitudinal one.} The spin transparency determines how effectively a given perturbation generates a spin torque. More explicitly, denoting the spin transparency for the interface by $T$ and assuming a spin current injection $\vec{j}_{s,\rm in}$, the resulting spin torque is determined by $T\vec{j}_{s, \rm in}$, not $\vec{j}_{s, \rm in}$ itself, because of the effects of spin diffusion in bulk and the reflection at boundaries. In this sense, the spin transparency can be understood by the absorption efficiency of a transverse spin current at the given interface. As one can expect, the spin transparency depends on the spin mixing conductance of the interface and the properties of the normal metal [see Eq.~(\ref{Eq:conventional T}) for an explicit expression]. Note that, for a trilayer system consisting of a thick film where the two interfaces hardly communicate with each other [Fig.~\ref{Fig:diff}(a)], the transparency at an interface is not affected by the properties of the normal metal at the other side.

However, recent interest of researches in magnetism has moved to ultrathin magnetic films, which not only allow high density spintronic applications but also result in much richer physics originating from broken symmetry such as spin-orbit torque (SOT)~\cite{Miron2011,Liu2012}, the Dzyaloshinskii-Moriya interaction~\cite{Dzaloshinsky1958,Moriya1960,Fert1980}, and other chiral phenomena~\cite{Schulz2012,Kim2012,Kim2013,Je2013,Yin2016,Moon2013}. The typical order of magnitudes of thicknesses of ferromagnetic layers under consideration is a few \AA\cite{Miron2011a,Liu2012a,Woo2016,Je2013,Emori2013,Ryu2013,Yin2016}, which cannot be assumed to be sufficiently larger than the spin dephasing length. Moreover, there are recent experimental reports on a ferrimagnetic multilayer with an extremely long spin coherence length $>10~\mathrm{nm}$~\cite{Yu2019} and a direct experimental evidence that the two interfaces of an ultrathin ferromagnet is no longer independent~\cite{Qiu2016}. Therefore, to correctly analyze the magnetic multilayers of contemporary research interest, it is desirable to construct a formal theory taking into account a finite dephasing rate of transverse spins and the resulting communication between the two interfaces of a magnetic layer.

For this purpose, it is necessary to examine the transport of a transverse spin current \emph{through} a ferromagnetic layer. To do this, we adopt the concept of the transmitted mixing conductance $G_T^\uda$ suggested by previous works~\cite{Qiu2016,Kovalev2006,Zwierzycki2005}. As visualized in Fig.~\ref{Fig:G_t}(a), the transmitted mixing conductance is the transmission counterpart of the conventional (reflected) spin mixing conductance (See Appendix~\ref{Sec(A):derivation G_T} for mathematical details). Therefore, it is a suitable concept for describing the intercommunication of the two interfaces of the ferromagnetic layer. However, its application to the drift-diffusion formalism has been very limited (to our knowledge, only to Ref.~\cite{Qiu2016}). Even in the previous attempt, they consider a particular limit where physical insight is more easily obtained, but the general solution is too complicated to go beyond such a simple case. In this paper, we introduce an \emph{effective} spin transparency, which provides a clear physical generalization of the conventional spin transparency, as well as significantly simplifies the complicated solution of the drift-diffusion equation in the general case. With the help of this general formalism, we demonstrate that the enhanced spin-orbit torque is realizable even without a special type of the interface~\cite{Qiu2016} and that there arises a nonnegligible field-like spin-orbit torque even in the absence of the imaginary part of the spin mixing conductance. Thus, introducing the effective transparency in our formalism is a useful tool to study general consequences of spin transport near an ultrathin ferromagnet.

This paper is organized as follows. In Sec.~\ref{Sec:Form}, we briefly review the conventional drift-diffusion formalism and present the modified BC for ultrathin films. In Sec.~\ref{Sec:consequencies}, we solve the drift-diffusion equation and calculate various physical quantities such as SOT, the inverse spin Hall current, and the spin pumping effect. To express our result in simple forms, we introduce an effective spin transparency. In Sec.~\ref{Sec:summary}, we summarize the paper. Appendixes include mathematical information that is not crucial for the main flow of the paper.

\section{Formalism\label{Sec:Form}}

\subsection{Review of the conventional spin drift-diffusion formalism\label{Sec:conventional Form}}

In this section, we review the spin drift-diffusion formalism with the conventional BC for thick ferromagnetic film. We consider an arbitrary magnetic multilayer system. It is usually assumed that the mean free path is much shorter than the spin diffusion length~\cite{Valet1993}, then  the spin chemical potential and the spin current in the normal metal bulk satisfy the spin drift-diffusion equation. Taking notations in Ref.~\cite{Chen2013}, the set of equations reads\footnote{In this paper, the longitudinal part of the equation is ignored and it does not affect the calculation of spin torque and spin pumping at all.}
\begin{subequations}\label{Eq:DDE0}
	\begin{equation}
	\partial_z^2\vec{\mu}_s=\frac{\vec{\mu}_s}{\lambda_i^2},\label{Eq:mus0}
	\end{equation}
	where $\lambda_i$ is the spin diffusion length for each normal metal layer and $i$ is the index of the layer. Since $\lambda_i$ can be different for each layer, the nonequilibrium chemical potential $\vec{\mu}_s$ is defined piecewisely. The spin current flowing along $z$ in the normal metal bulk is given by the drift-diffusion current:
	\begin{equation}
	\vec{j}_s(z)=-\frac{\sigma_i}{2e}\partial_z\vec{\mu}_s,\label{Eq:js0}
	\end{equation}
	where $\sigma_i$ is the electrical conductivity of the normal metal $i$ and $e>0$ is the electron charge. In Eq.~(\ref{Eq:js0}), one may introduce additional term if there is another current source. A famous example is the spin Hall current that we introduce in Sec.~\ref{Sec:transparency}.
\end{subequations}

To obtain the full solution of $\vec{\mu}_s$ and $\vec{j}_s$, one should apply proper BCs at each of the interfaces between two layers. The form of the BC depends on the type of the interface. Suppose that there is an interface at $z=z_0$. (i) For an interface between a normal metal layer and the vacuum, $\vec{j}_s(z_0)=0$ should be satisfied. (ii) For an interface between two normal metal layers, $\vec{j}_s=G\Delta\vec{\mu}_s$ is satisfied where $G$ is the interface conductance of the interface and $\Delta\vec{\mu}_s$ is the spin chemical potential drop at the interface. (iii) For an interface between a normal metal and a ferromagnet, the BC is given by the circuit theory [Eq.~(\ref{Eq:conventional BC0})].

Now we explicitly apply this knowledge to the magnetic trilayer depicted in Fig.~\ref{Fig:diff} and construct our model. The magnetic trilayer consists of a ferromagnetic layer (FM) sandwiched by two normal metal layers (NM1 and NM2): NM1($[-d_1,0]$)/FM($[0,t]$)/NM2($[t,t+d_2]$), where $d_1$ and $d_2$ are the thicknesses of the normal metal layers and $t$ is the thickness of the ferromagnet. First of all, the equation in NM1 and NM2 is piecewisely given by Eq.~(\ref{Eq:DDE0}) where we denote NM1 and NM2 by $i=1,2$ respectively. For the boundaries with the vacuum,
\begin{equation}
\vec{j}_s(-d_1)=\vec{j}_s(t+d_2)=0\label{Eq:BC vacuum}
\end{equation}
should be satisfied. For the BC at $z=0$ and $z=t$ (interfaces between normal metals and the ferromagnet),\footnote{In Eq.~(\ref{Eq:conventional BC(a)}), the presence of the minus sign in front of $\vec{\mu}_s(0^-)$ is because $\Delta\vec{\mu}_s(0)=-\vec{\mu}_s(0^-)$.}
\begin{subequations}\label{Eq:conventional BC}
	\begin{align}
	e\vec{j}_s(0^-)&=\Re[-G_1^\uda \mathcal{M}\vec{\mu}_s(0^-)],\label{Eq:conventional BC(a)}\\
	e\vec{j}_s(t^+)&=\Re[G_2^\uda \mathcal{M}\vec{\mu}_s(t^+)],
	\end{align}
\end{subequations}
where $G_i^\uda$ is the spin mixing conductance of each interface ($i=1,2$), $\mathcal{M}$ is a linear operator defined by $\mathcal{M}\vec{v}=\vec{m}\times(\vec{m}\times\vec{v}-i\vec{v})$ for a three-dimensional vector $\vec{v}$, which allows compactly expressing the two terms in Eq.~(\ref{Eq:conventional BC0}) by a single term. If there is time-varying magnetization, the spin pumping yields additional terms in the BC~\cite{Tserkovnyak2002,Lee2013} as we consider in Sec.~\ref{Sec:SP}.

To calculate the SOT per unit area, we use the angular momentum conservation. Note that $\vec{j}_s(0^-)$ is the incoming angular momentum to the ferromagnet and $\vec{j}_s(t^+)$ is the outgoing angular momentum from the ferromagnet. Therefore, the angular momentum absorbed by the ferromagnet is given by $\vec{j}_s(0^-)-\vec{j}_s(t^+)$. Considering the conversion factor from the electrical current and the spin angular momentum, the SOT per unit area $\vec{\tau}$ is given by 
\begin{equation}\label{Eq:SOT def}
\vec{\tau}=\frac{\hbar}{2e}[\vec{j}_s(0^-)-\vec{j}_s(t^+)].
\end{equation}

\subsection{Modified BC by the transmitted mixing conductance\label{Sec:TMC}}

\begin{figure}
	\includegraphics[width=8.6cm]{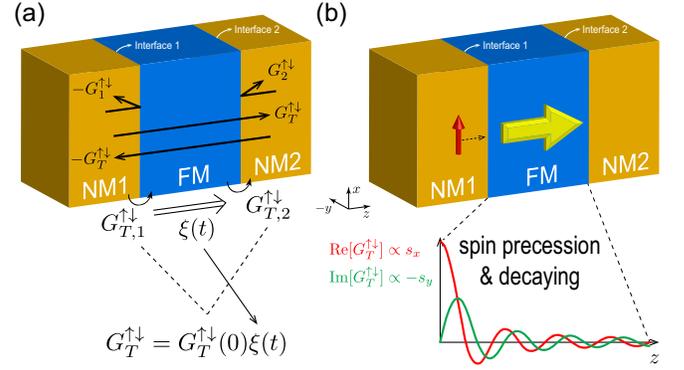}
	\caption{(a) (top) Illustration of the definitions of the conventional ($G_i^\uda$) and the transmitted ($G_T^\uda$) mixing conductances. (bottom) The transmitted mixing conductance is determined by the product of the interface discontinuity at each interface ($G_{T,1}^\uda$ and $G_{T,2}^\uda$) and the bulk contribution $\xi(t)$, which refers to the spin dephasing and is thickness dependent. (b) The decaying and oscillatory nature of the spin dephasing, as a result of the rotation of the transverse spin around the magnetization. The plot is generated for a real $G_T^\uda(0)$.
		\label{Fig:G_t}}
\end{figure}

In this section, we modify the formalism in Sec.~\ref{Sec:conventional Form} to take an ultrathin film into account. When the thickness of the magnet $t$ is not much larger than the spin dephasing length, the spin chemical potential at the interface~1 can generates the spin current at the interface 2 (and vice versa) [Fig.~\ref{Fig:diff}(b)]. In this case, it is necessary to introduce another conductance $G_T^\uda$, called the \emph{transmitted} mixing conductance~\cite{Qiu2016} and whose properties are discussed below. As illustrated in the top part of Fig.~\ref{Fig:G_t}(a), the transmitted mixing conductance connects $\vec{\mu}_s(t^+)$ and $\vec{j}_s(0^-)$ (and vice versa), giving the following modified BC.
\begin{subequations}\label{Eq:modified BC}
\begin{align}
e\vec{j}_s(0^-)&=\Re[-G_1^\uda\mathcal{M}\vec{\mu}_s(0^-)+G_T^\uda\mathcal{M}\vec{\mu}_s(t^+)],\label{Eq:modified BC(a)}\\
e\vec{j}_s(t^+)&=\Re[-G_T^\uda\mathcal{M}\vec{\mu}_s(0^-)+G_2^\uda\mathcal{M}\vec{\mu}_s(t^+)].\label{Eq:modified BC(b)}
\end{align}
\end{subequations}
Equation~(\ref{Eq:modified BC}) gives a simple extension of the conventional BC [Eq.~(\ref{Eq:conventional BC})] to allow for a finite dephasing rate. The formal derivation of Eq.~(\ref{Eq:modified BC}) is presented in Appendix~\ref{Sec(A):derivation G_T}.

There are three physical processes behind the transmitted mixing conductance [bottom part of Fig.~\ref{Fig:G_t}(a)]. First, when a transverse spin is injected to and passing through the interface~1, there arise the interfacial spin filtering and the interfacial spin rotation, which make the spin current discontinuous at the interface $\vec{j}_s(0+)\ne\vec{j}_s(0^-)$. The details of the interfacial spin filtering and rotation are substantially discussed in Ref.~\cite{Stiles2002}. In Fig.~\ref{Fig:G_t}(a), we denote this process by $G_{T,1}^\uda$. The second process is the spin dephasing in the bulk of the magnetic layer. In this work, the spin dephasing is characterized by a thickness-dependent function $\xi(t)$, whose features for various materials are discussed below. The third process is the additional spin filtering and rotation at the interface~2 denoted by $G_{T,2}^\uda$ in Fig.~\ref{Fig:G_t}(a). Now, we may write the transmitted mixing conductance by the following form.
\begin{equation}
G_T^\uda(t)=G_T^\uda(0)\xi(t),\label{Eq:decomposition}
\end{equation}
where $t$ is the thickness of the magnet. $G_T^\uda(0)$ is the interfacial contribution determined by $G_1^\uda$ and $G_2^\uda$. $\xi(t)$ is the bulk contribution and satisfies $\xi(0)=1$ (no spin dephasing) and $\xi(\infty)=0$ (perfect spin dephasing).

\begin{subequations}
Inside the magnetic layer, the transverse spin current decays rapidly over the spin dephasing length whose mechanisms determine the properties of $\xi(t)$. There are multiple origins of the spin dephasing; coherent and incoherent scatterings. For instance, in ferromagnetic metals, coherent spin oscillation of a number of electrons in the Fermi sea with different momenta is one of the main origins of the spin dephasing. In this case, $\xi(t)$ is oscillatory and decaying approximately in the form of~\cite{Stiles2002}
\begin{equation}
\xi(t)=j_0(\pi t/\lambda_{\rm dp})+ij_1(\pi t/\lambda_{\rm dp}),\label{Eq:xi for metal}
\end{equation}
where $\lambda_{\rm dp}=\pi/(k_F^\uparrow-k_F^\downarrow)$ is the spin dephasing length, $k_F^\sigma$ is the Fermi wave vector for spin $\sigma$, and $j_n$ is the spherical Bessel function: $j_0(x)=(1/x)\sin x$ and $j_1=-j_0'(x)$. Indeed, this analytic form is valid for large $t$ limit (under the stationary phase approximation), but numerical calculations for $G_T$~\cite{Stiles2002,Zwierzycki2005} implies that this approximation works reasonably well in the aspect of qualitative understanding. For 3$d$ transition metals, $\lambda_{\rm dp}$ is on the order of a nanometer~\cite{Zhang2004}, thus $t/\lambda_{\rm dp}$ is on the order of one for an ultrathin magnet. The physical interpretation of the oscillatory and decaying nature of $\xi$ is the precession of the incident spin around the magnetization, as illustrated in Fig.~\ref{Fig:G_t}(b). In the case that there are incoherent scattering sources, an additional exponential factor can be introduced, but it results in quantitative corrections only.

Although we focus on the ferromagnetic metal case [Eq.~(\ref{Eq:xi for metal})] in explicit numerical computations below, we discuss the form of $\xi(t)$ for other systems as well. In most cases, an exponentially decaying $\xi(t)=e^{-t/\lambda_{\rm dp}}$ is relevant. For example, in ferromagnetic insulators, the spin current is injected as magnon excitations, which decay over the spin-wave attenuation length or the magnon diffusion length~\cite{Cornelissen2016}. For systems with an extremely large coherence length~\cite{Yu2019}, the spin diffusion length would be the relevant length scale. For a magnet showing spin superfluidity~\cite{Takei2014}, the spin current decays algebraically rather than exponentially.

Two remarks are in order. First, although we only consider trilayers for writing Eq.~(\ref{Eq:modified BC}), generalization of our theory to multilayer is straightforward. This is because normal metals are typically in the regime where the drift-diffusion equation is valid. Thus, one can write down the drift-diffusion equation in each layer and apply the modified BC [Eq.~(\ref{Eq:modified BC})] for all embedded magnetic layers. Second, consideration of the effects of interfacial spin-orbit coupling~\cite{Amin2016,Amin2016a,Kim2017} goes beyond the scope of this paper. An additional consideration of the interface-generated spin current~\cite{Amin2018} would be a way to generalize the formalism.
\end{subequations}

\section{Physical consequences\label{Sec:consequencies}}

\subsection{Transparency for injecting a spin Hall current\label{Sec:transparency}}

\begin{figure}
	\includegraphics[width=8.6cm]{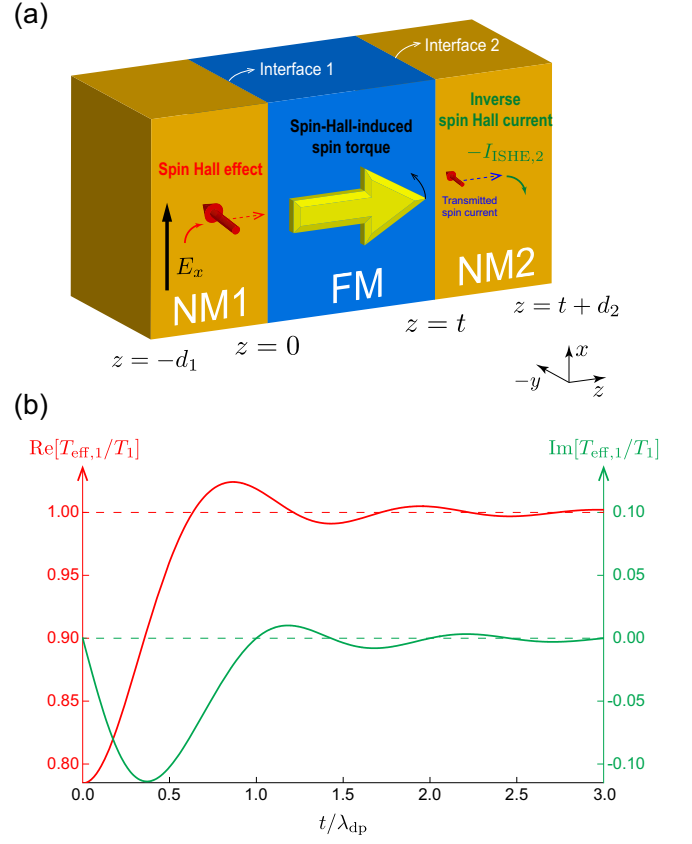}
	\caption{(a) Illustration of the generation of spin-Hall-effect-induced SOT. When an electric field is applied along the $x$ direction in NM1, a spin Hall current (red) is generated and gives rise to a torque (black). If the spin dephasing is not perfect, the transmitted current (blue) generates additional inverse spin Hall current in NM2 (green), which is absent for a thick ferromagnet. (b) Plot of real (red) and imaginary (green) parts of the effective spin transparency [Eq.~(\ref{Eq:Teff})]. Here, the real and imaginary parts represent, respectively, the change of the spin torque efficiency and  the generation of a field-like SOT (even for $\Im G_i^\uda=0$). Here we use $G_i^\uda=(e^2/h)\times 100~\mathrm{nm}^{-2}$, $\sigma_i=(15~\mu\Omega~\mathrm{cm})^{-1}$, and $\lambda_i=1.4~\mathrm{nm}$ by referring to Ref.~\cite{Zhang2015} and take $d_i=1~\mathrm{nm}$ and $G_T^\uda(0)=0.5$.
	\label{Fig:spin Hall injection}}
\end{figure}

One of the most frequently performed experiments with ultrathin ferromagnets is injecting a spin Hall current to a ferromagnet to generate SOT~\cite{Liu2012,Liu2012a,Zhang2015,Liu2011}. Figure~\ref{Fig:spin Hall injection}(a) shows the experimental situation under our consideration. When an electrical current is applied along the $x$ direction in NM1, the spin Hall effect~\cite{Sinova2015} generates a torque to FM. The injection efficiency is determined by the spin transparency proposed in Ref.~\cite{Zhang2015}. As demonstrated in the previous paper and discussed in Sec.~\ref{Sec:introduction}, if the ferromagnet is thick enough, the two interfaces do not communicate with each other, thus the transparency of the interface 1 is solely determined by the properties of NM1 and the spin mixing conductance of the interface 1. However, if the spin dephasing is not perfect, the situation is no longer as simple as the previous result.

To calculate the spin Hall effect contribution, we add one more current source in NM1. The current in Eq.~(\ref{Eq:js0}) is modified as
\begin{equation}\label{Eq:jsSHE}
\vec{j}_s(z)=\begin{cases}\displaystyle
-\frac{\sigma_1}{2e}\partial_z\vec{\mu}_s-\sigma_{\rm SH,1}E_x\vhat{y}&\text{in NM1}\\[10pt]
\displaystyle
-\frac{\sigma_2}{2e}\partial_z\vec{\mu}_s&\text{in NM2},
\end{cases}
\end{equation}
where $\sigma_{\rm SH,\it i}$ is the spin Hall conductivity of each normal metal, and $E_x$ is the applied electric field in NM1 along the $x$ direction.\footnote{The spin Hall current contribution is absent in NM2 since no electric field is applied there. Consideration of an additional electric field applied in NM2 is very straightforward because the drift-diffusion equation is linear: calculating the consequences of electric field applied in each layer \textit{separately}, and simply adding up the two results.} 

Now one can obtain the spatial profile of $\vec{j}_s$ by solving Eqs.~(\ref{Eq:mus0}) and (\ref{Eq:jsSHE}) with the BCs in Eqs.~(\ref{Eq:BC vacuum}) and Eq.~(\ref{Eq:modified BC}). From the explicit solution available in Appendix~\ref{Sec(A):full solution}, one can use Eq.~(\ref{Eq:SOT def}) to obtain the SOT per unit area:
\begin{subequations}\label{Eq:transparencies}
\begin{align}
\vec{\tau}&=\frac{\hbar}{2e}\sigma_{\rm SH,1}E_x\Re[T_{\rm eff,1}\mathcal{M}\vhat{y}],\label{Eq:SOT}\\
T_{\rm eff,1}&=T_1\frac{1-\tilde{G}_T^{\uda}(t)^2T_2/T_2'}{1-\tilde{G}_T^{\uda}(t)^2T_{12}},\label{Eq:Teff}
\end{align}
where $T_{\rm eff,1}$ is the \emph{effective} spin transparency for the interface~1, $\tilde{G}_T^\uda(t)=G_T^\uda(t)/\sqrt{G_1^\uda G_2^\uda}$ is the normalized transmitted mixing conductance (dimensionless), $T_i$ is the conventional spin transparency for the interface $i=1,2$,
\begin{equation}
T_i=\frac{G_i^\uda\tanh\frac{d_i}{2\lambda_i}}{G_i^\uda\coth\frac{d_i}{\lambda_i}+\frac{\sigma_i}{2\lambda_i}},\label{Eq:conventional T}
\end{equation}
and the other transparency-like quantities are given by
\begin{align}
T_2'&=\frac{G_T^\uda\tanh\frac{d_2}{2\lambda_2}}{G_T^\uda(t)\coth\frac{d_2}{\lambda_2}+\frac{\sigma_2}{2\lambda_2}},\\
T_{12}&=T_1T_2\frac{\coth\frac{d_1}{\lambda_1}\coth\frac{d_2}{\lambda_2}}{\tanh\frac{d_1}{2\lambda_1}\tanh\frac{d_2}{2\lambda_2}}.
\end{align}
\end{subequations}
The effective spin transparency is the central result of this paper.
In the expression of $T_{\rm eff}$, $\xi(t)$ appears indirectly through $\tilde{G}^\uda(t)$. Note that Eq.~(\ref{Eq:Teff}) restores the previously reported result $T_{\rm eff}=T_1$~\cite{Zhang2015} for $t\to\infty$ where $\xi(t)\to 0$. For later purpose, we also define $T_{\rm eff, 2}$ and $T_1'$ by the same way as Eq.~(\ref{Eq:transparencies}) except the exchange between subscripts $1$ and $2$.

For simplicity of analysis, we assume that $G_i^\uda$ are positive real numbers and $|G_T^\uda|<G_i^\uda$ as considered in most experimental situations~\cite{Zwierzycki2005}. One can easily prove that $|T_{\rm eff}|$ is always smaller than $T_1$ (thus SOT cannot be enhanced) if $G_T^\uda$ is a positive real number. To mathematically show this, we use Eq.~(\ref{Eq:Teff}) and verify that $|T_{\rm eff,1}|<T_1$ holds if and only if $T_2/T_2'>T_{12}$ (See Appendix~\ref{Sec(A):condition1} for proof). In addition, it is also easy to show that $T_{12}<T_2/T_2'$ if $G_T^\uda$ is positive and real (See Appendix~\ref{Sec(A):condition2} for proof). This gives $|T_{\rm eff,1}|<T_1$, concluding the proof. Therefore, SOT is unlikely to be enhanced for a positive and real $G_T^\uda$.

However, in more general cases that $G_T^\uda$ is not a positive real number, SOT can be enhanced. For metallic cases described by Eq.~(\ref{Eq:xi for metal}), $\Im G_T^\uda$ is on the same order of magnitude as $\Re G_T^\uda$, thus $G_T^\uda$ cannot be assumed to be positive and real. Furthermore, Eq.~(\ref{Eq:xi for metal}) implies that $G_T^\uda$ can even be a negative real number, as also demonstrated in Refs.~\cite{Zwierzycki2005,Kovalev2006,Balaz2016}. For this case, $T_2'/T_2<0$, thus it is always smaller than $T_{12}>0$. Thus SOT can be enhanced for a negative $G_T^\uda$. More explicitly, we take Eq.~(\ref{Eq:xi for metal}) for $\xi(t)$ and plot $T_{\rm eff,1}/T_1$ as a function of $t$ in Fig~\ref{Fig:spin Hall injection}(b). It clearly shows that, for some regions ($t\lesssim\lambda_{\rm dp}$), the spin torque can be enhanced ($\Re[T_{\rm eff,1}]>T_1$) and there arises a nonnegligible field-like component of SOT ($\Im[T_{\rm eff,1}]\not\approx0$) even for $\Im G_i^\uda=0$, which makes a qualitative difference from thick film cases.

The enhancement of spin torque can be understood by Fig.~\ref{Fig:diff}(b)~\cite{Kovalev2006}. For $\lambda_{\rm dp}<t<2\lambda_{\rm dp}$, Eq.~(\ref{Eq:xi for metal}) has a negative real part, thus $s_x(z=t)$ in Fig.~\ref{Fig:diff}(b) can be negative. Thus, the angular momentum transfer to the ferromagnet [$s_x(z=0)-s_x(z=t)$] is larger than $s_x(z=0)$. A recent experiment~\cite{Qiu2016} also suggests that the negativity of $G_T^\uda$ may enhance the SOT. In that experiment, the spin flip precisely at $z=t$ may result in $s_x(z=t)$ being negative. This is an \textit{interfacial} contribution ($\Re[G_T^\uda(0)]<0$ in our convention), while the enhanced spin transparency in Fig.~\ref{Fig:spin Hall injection}(b) originates from the \textit{bulk} contribution ($\Re[\xi(t)]<0$) not requiring such a special interface.

\subsection{Inverse spin Hall effect from NM2\label{Sec:ISHE}}

\begin{figure}
	\includegraphics[width=8.6cm]{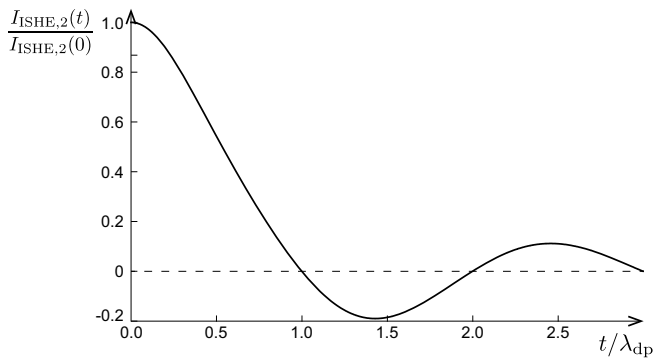}
	\caption{Thickness dependence of $I_{\rm ISHE,2}$ [Eq.~(\ref{Eq:ISHE2})]. We use the same parameter as Fig.~\ref{Fig:spin Hall injection}.
		\label{Fig:ISHE2}}
\end{figure}

One of physical consequences that are absent for $G_T^\uda=0$ but present for $G_T^\uda\ne0$ is the inverse spin Hall current in NM2. As depicted in Fig.~\ref{Fig:spin Hall injection}(a), when an electric field is applied in NM1, the injected spin Hall current from NM1 can reach $z=t^+$ (blue) since the dephasing in the ferromagnetic bulk is not perfect. The nonzero spin current at $z=t^+$ may give rise to an inverse spin Hall current along $x$ in NM2 (green). To calculate the resulting charge current along $x$, we assume that $\vec{m}$ is perpendicular to the injected spin current $\sigma_{\rm SH,1}E_x\vhat{y}$ since transport of a longitudinal spin in the ferromagnet is beyond the scope of this paper. The total inverse spin Hall current in NM2 is given by $I_{\rm ISHE,2}=W\int_t^{t^+d_2}\theta_{\rm SH,2}\vhat{y}\cdot\vec{j}_s(z)dz$, where $W$ is the width of the wire and $\theta_{\rm SH,\it i}=\sigma_{\rm SH,\it i}/\sigma_i$ is the spin Hall angle, and $\sigma_{\rm SH,2}$ is the spin Hall conductivity of NM2. Using the solution in Appendix~\ref{Sec(A):full solution} for $\vec{j}_s(z)$, we obtain
\begin{equation}\label{Eq:ISHE2}
I_{\rm ISHE,2}=-\frac{\sigma_{\rm SH,1}\sigma_{\rm SH,2}E_xW}{2}\Re\left[\frac{G_T^\uda }{G_1^\uda G_2^\uda}\frac{T_{\rm eff,1}T_2 T_2'}{T_2'-\tilde{G}_T^{\uda2} T_2} \right].
\end{equation}
In Fig.~\ref{Fig:ISHE2}, We plot $I_{\rm ISHE,2}$ as a function of thickness with using the ansatz Eq.~(\ref{Eq:xi for metal}). It changes the sign at $t=\lambda_{\rm dp}$, since the damping-like component of the transmitted spin changes its sign at this point. There are two remarks. First, from the expressions in Eq.~(\ref{Eq:transparencies}), one can prove that Eq.~(\ref{Eq:ISHE2}) is symmetric under the exchange $1\leftrightarrow2$, as guaranteed by the Onsager reciprocity. Second, when an electric field is applied along NM1, a shunting current flowing through NM2 can affect the measurement of $I_{\rm ISHE,2}$. To eliminate such contributions, one may use a charge insulator as the ferromagnet or an insertion layer.

\begin{figure}[b]
	\includegraphics[width=7.5cm]{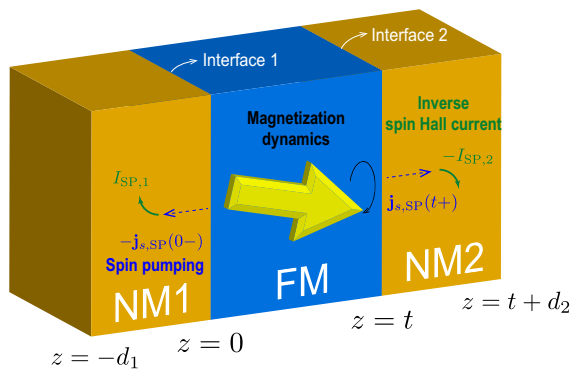}
	\caption{Geometry of the spin pumping calculation considered in Eq.~(\ref{Eq:BC_SP}). In the presence of magnetization dynamics (black), the spin pumping currents are generated at both interfaces [denoted by $\vec{j}_{s, \rm SP}(0^-)$ and $\vec{j}_{s, \rm SP}(t^+)$, blue], which further generates the inverse spin Hall current in each layer (denoted by $I_{\rm SP,\it i}$).
		\label{Fig:spin pumping}}
\end{figure}

\subsection{Spin pumping\label{Sec:SP}}
Spin pumping~\cite{Tserkobnyak2005,Tserkovnyak2002} is another physical phenomenon in which the mixing conductances play an important role. It is frequently used for measuring the spin transparency~\cite{Lee2018}, the spin Hall angle~\cite{RojasSanchez2014,Weiler2014,Wang2014}, and the spin diffusion length~\cite{Wang2014,Zhang2013}. Here, we examine the effect of a nonzero $G_T^\uda$ on spin pumping for the geometry depicted in Fig.~\ref{Fig:spin pumping}. In the presence of magnetization dynamics $\partial_t\vec{m}\ne0$, angular momentum is pumped to both normal metals, as so-called the spin pumping currents (blue). These pumped currents generate measurable inverse spin Hall currents along the $x$ direction in each normal metal, which are denoted by $I_{\rm SP,\it i}$ (green). To calculate these currents, one needs to take into account the spin pumping currents as additional BCs. Taking the theory of spin pumping~\cite{Tserkovnyak2002}, we add
\begin{subequations}\label{Eq:BC_SP}
\begin{align}
e\vec{j}_{s,\rm SP}(0^-)&=\frac{\hbar}{2}\Re[(G_T^\uda-G_1^\uda)\mathcal{M}(\vec{m}\times\partial_t\vec{m})],\\
e\vec{j}_{s,\rm SP}(t^+)&=\frac{\hbar}{2}\Re[(G_2^\uda-G_T^\uda)\mathcal{M}(\vec{m}\times\partial_t\vec{m})],
\end{align}
\end{subequations}
to Eqs.~(\ref{Eq:modified BC(a)}) and Eq.~(\ref{Eq:modified BC(b)}), respectively. Solving the same drift-diffusion equation [Eqs.~(\ref{Eq:mus0}) and (\ref{Eq:jsSHE})] without the external electric field ($E_x=0$), one obtains the spin current profile $\vec{j}_s(z)$ and the resulting inverse spin Hall currents in NM1 and NM2 given by $I_{\rm SP,\it i}=W\int_{\rm NM\it i} \theta_{\rm SH,\it i}\vhat{y}\cdot\vec{j}_s(z)dz$. After some algebra,
\begin{equation}
I_{\rm SP,\it i}=(-1)^i\frac{W\hbar}{4e}\sigma_{\rm SH,\it i}(\vec{m}\times\partial_t\vec{m})\cdot\Re[T_{\rm eff,\it i}^*\mathcal{M}\vhat{y}].
\end{equation}
The appearance of the same $T_{\rm eff, \it i}$ is understood by the Onsager reciprocity of spin pumping and spin torque. The factor $(-1)^i$ is also understandable by Fig.~\ref{Fig:spin pumping} where the directions of the spin pumping currents to NM1 and NM2 are opposite. The inverse spin Hall measurement of the spin pumping effect can give $T_{\rm eff, \it i}$ separately.

However, the enhanced Gilbert damping~\cite{Tserkovnyak2002} from the spin pumping effect requires more carefulness. This is because the Gilbert damping enhancement $\Delta\alpha_{\rm SP}$ is not strictly given by the Onsager reciprocity when the system consists of multiple sources (interfaces 1 and 2) of angular momentum pumping. To calculate $\Delta\alpha_{\rm SP}$, we calculate the total angular momentum transfer per unit area as $\vec{\tau}=(\hbar/2e)[\vec{j}_s(0^-)-\vec{j}_s(t^+)]$ and project $\vec{\tau}$ to $\vec{m}\times\partial_t\vec{m}$ to obtain its coefficient. Neglecting the renormalization of the gyromagnetic ratio~\cite{Tserkovnyak2002}, we obtain
\begin{equation}
\Delta\alpha_{\rm SP}=\frac{\gamma\hbar^2}{8M_se^2t}\Re\left[\sum_{i=1,2}\frac{\sigma_i}{\lambda_i}T_{\rm eff,\it i}\coth\frac{d_i}{2\lambda_i}\right],
\end{equation}
where $\gamma$ is the gyromagnetic ratio and $M_s$ is the saturation magnetization. Note that $\Delta\alpha_{\rm SP}$ is given by the sum of $T_{\rm eff}$ for each interface with some weighting factors. Since the weighting factors [$(\sigma_i/\lambda_i)\coth(d_i/2\lambda_i)$] for each interface are not identical, extracting $T_{\rm eff,\it i}$ from measurement of $\Delta\alpha_{\rm SP}$ requires more experimental information.

\section{Summary\label{Sec:summary}}
In summary, we consider the effects of a nonzero transmitted mixing conductance in the drift-diffusion formalism to allow for the finite rate of the spin dephasing in an ultrathin ferromagnetic whose thickness is not much larger than the spin dephasing length. Solving the drift-diffusion equation with a modified BC, we demonstrate that spin torque can be enhanced in thin films, because of rotation of an injected spin current in ferromagnetic metals. Moreover, a nonnegligible field-like SOT can arise even in the absence of the imaginary part of the conventional spin mixing conductance. We demonstrate these by simply introducing an effective spin transparency, which also appears in the expression of the spin pumping current and the resulting Gilbert damping enhancement. The effective spin transparency obtained here provides a simple and straightforward extension of the conventional BC of the drift-diffusion formalism.

\begin{acknowledgments}
The author acknowledges B.~C.~Min and O.~J.~Lee for motivating this work, D.~S.~Han for discussions, and K.~J.~Lee for critical reading of the manuscript. This work was financially supported by the KIST Institutional Program, the National Research Council of Science \& Technology (NST) (Grant No. CAP-16-01-KIST), and the German Research Foundation (DFG) (No. SI 1720/2-1).
\end{acknowledgments}

\begin{appendix}
\section{Derivation of the transmitted mixing conductance\label{Sec(A):derivation G_T}}

To derive Eq.~(\ref{Eq:modified BC}), it is required to extend the circuit theory~\cite{Brataas2000,Brataas2001} to multiple interfaces. This is done in Sec 7.1 of Ref.~\cite{Brataas2006}. According to the theory, the current in NM1 side reads
\begin{equation}
\hat{i}^{(1)}=-\frac{1}{2}\sum_{nm}[\hat{t}_{nm}\hat{\mu}^{(2)}\hat{t}_{nm}^\dagger-\hat{\mu}^{(1)}+\hat{r}_{nm}\hat{\mu}^{(1)}\hat{r}_{nm}^\dagger],
\end{equation}
where $\hat{t}_{nm}$ and $\hat{r}_{nm}$ are the transmission and reflection matrices for the transverse mode [denoted by $(n,m)$] incident from NM1 ($z<0$) and $\hat{\cdot}$ is the $2\times2$ matrix in the Pauli spin space. The scattering matrices are defined by the scattering process over \emph{the entire ferromagnet} consisting of two interfaces and bulk (not a single interface) (see Fig.~1 of Ref.~\cite{Tserkovnyak2002} for a similar example). Compared to Ref.~\cite{Brataas2006}, an additional minus sign appears in our notation, since it is the current to the $-z$ direction. Disregarding the charge degree of freedom, the relations between $\hat{\mu}^{(i)}$ and $\vec{\mu}_s$ in our theory are given by $\hat{\mu}^{(1)}=\vec{\mu}_s(0^-)\cdot\vec{\sigma}$ and $\hat{\mu}^{(2)}=\vec{\mu}_s(t^+)\cdot\vec{\sigma}$ where $\vec{\sigma}$ is the Pauli matrix. Following the procedure in Ref.~\cite{Brataas2001}, we disregard the spin-flip process in the contacts and write down the reflection and transmission matrices as
\begin{equation}
\hat{r}_{nm}=\sum_{s=\uparrow,\downarrow}\hat{u}_sr_{nm}^s,~\hat{t}_{nm}=\sum_s\hat{u}_st_{nm}^s,
\end{equation}
where $\hat{u}^{\uparrow/\downarrow}=(1\pm\vec{\sigma}\cdot\vec{m})/2$ is the spin-projection matrix. In this regime, the current matrix can be expressed in terms of $\vec{\mu}_s(0^-)$, $\vec{\mu}_s(t^+)$, $r_{nm}^s$, and $t_{nm}^s$. Then, the current $\vec{j}_s(0^-)$ is proportional to the transverse component of $\Tr[\hat{i}^{(1)}\vec{\sigma}]/2$.

After some algebra, we obtain
\begin{align}
\frac{1}{2}\Tr[\hat{i}^{(1)}\vec{\sigma}]&=g_e\{\vec{m}\cdot\vec{\mu_s}(0^-)-\vec{m}\cdot\vec{\mu_s}(t^+)\}\vec{m}\nonumber\\
&\quad+\Re[-g_r^\uda\mathcal{M}\vec{\mu}_s(0^-)+g_t\mathcal{M}\vec{\mu}_s(t^+)],\label{Eq(A):i}
\end{align}
where $g_e=(1/4)\sum_{nm,s}|t_{nm}^s|^2$ corresponds to the longitudinal transport, $g_r=(1/2)(M-\sum_{nm}r_{nm}^\uparrow r_{nm}^{\downarrow*})$ corresponds to the conventional mixing conductance, and $g_t=(1/2)\sum_{nm} t_{nm}^\uparrow t_{nm}^{\downarrow*}$ corresponds to the transmitted mixing conductance (See Fig.~\ref{Fig:G_t}). Here $M$ is the number of transverse modes. Taking only transverse part [second term in Eq.~(\ref{Eq(A):i})] with introducing a proportionality constant connecting $\Tr[\hat{i}^{(1)}\vec{\sigma}]/2$ and $\vec{j}_s$ gives Eq.~(\ref{Eq:modified BC(a)}).

Equation~(\ref{Eq:modified BC(b)}) can be obtained by a similar way. Note that the Onsager reciprocity~\cite{Hals2010} guarantees that the transmitted conductances in Eqs.~(\ref{Eq:modified BC(a)}) and (\ref{Eq:modified BC(b)}) are identical.

\begin{widetext}
\section{Explicit solution of the spin drift-diffusion equation for spin Hall injection\label{Sec(A):full solution}}
After solving Eqs.~(\ref{Eq:mus0}) and (\ref{Eq:jsSHE}) with the BCs in Eqs.~(\ref{Eq:BC vacuum}) and Eq.~(\ref{Eq:modified BC}), one obtains the chemical potential,
\begin{subequations}
	\begin{equation}
		\vec{\mu}_s(z)=\begin{cases}\displaystyle
		2eE_x\lambda_1\theta_{\rm SH,1}\Re\left[\frac{\frac{T_1}{G_1}\left(G_1\frac{\sinh\frac{z}{2\lambda_1}}{\sinh\frac{d_1}{2\lambda_1}}\frac{\cosh\frac{z}{2\lambda_1}}{\sinh\frac{d_1}{2\lambda_1}}+\frac{\sigma_1}{2\lambda_1}\frac{\sinh\frac{z+(d_1/2)}{\lambda_1}}{\sinh\frac{d_1}{2\lambda_1}}\right)-\tilde{G}_T^{\uda2} T_{12}\frac{\sinh\frac{z}{\lambda_1}}{\cosh\frac{d_1}{\lambda_1}}}{1-\tilde{G}_T^{\uda2}T_{12}}\mathcal{M}\vhat{y}\right]&\text{in NM1},\\[10pt]
		\displaystyle
		\frac{1}{2}e\sigma_{\rm SH,1}E_x \cosh\frac{z-(t+d_2)}{\lambda_2}\csch^2\frac{d_2}{2\lambda_2}\Re\left[\frac{G_T^\uda T_1T_2}{G_1^\uda G_2^\uda-G_T^{\uda2}T_{12}}\mathcal{M}\vhat{y}\right]&\text{in NM2},
		\end{cases}
	\end{equation}
and the current,
\begin{equation}
\vec{j}_s(z)=\begin{cases}\displaystyle
\sigma_{\rm SH,1}E_x\Re\left[\left(1-\frac{\frac{T_1}{G_1}\left(G_1\frac{\cosh\frac{z}{\lambda_1}}{2\sinh^2\frac{d_1}{2\lambda_1}}+\frac{\sigma_1}{2\lambda_1}\frac{\cosh\frac{z+(d_1/2)}{\lambda_1}}{\sinh\frac{d_1}{2\lambda_1}}\right)-\tilde{G}_T^{\uda2} T_{12}\frac{\cosh\frac{z}{\lambda_1}}{\cosh\frac{d_1}{\lambda_1}}}{1-\tilde{G}_T^{\uda2}T_{12}}\right)\mathcal{M}\vhat{y}\right]&\text{in NM1},\\[10pt]
\displaystyle
-\frac{1}{4\lambda_2}\sigma_{\rm SH,1}\sigma_2E_x \sinh\frac{z-(t+d_2)}{\lambda_2}\csch^2\frac{d_2}{2\lambda_2}\Re\left[\frac{G_T^\uda T_1T_2}{G_1^\uda G_2^\uda-G_T^{\uda2}T_{12}}\mathcal{M}\vhat{y}\right]&\text{in NM2}.
\end{cases}
\end{equation}
\end{subequations}
\end{widetext}

\section{Condition for $|T_{\rm eff,1}|<T_1$ for a real $G_T^\uda$\label{Sec(A):condition1}}
Provided that all the mixing conductances are real, all transparencies defined in Eq.~(\ref{Eq:transparencies}) are real. We first look at the denominator of
\begin{equation}
\frac{T_{\rm eff,1}}{T_1}=\frac{1-\tilde{G}_T^{\uda2}(T_2/T_2')}{1-\tilde{G}_T^{\uda2}T_{12}}.
\end{equation}
Note that we assume $|\tilde{G}_T^\uda|<1$ and $0<T_{12}<1$ (see Appendix~\ref{Sec(A):condition2}), we obtain
\begin{equation}
0<\tilde{G}_T^{\uda2}T_{12}<1,
\end{equation}
implying that that the denominator is positive.

Then we look at the numerator. By noting that
\begin{align}
\left|\tilde{G}_T^{\uda2}\frac{T_2}{T_2'}\right|&
<\left|\frac{G_T^\uda\coth\frac{d_2}{\lambda_2}+\frac{\sigma_2}{2\lambda_2}}{G_2^\uda\coth\frac{d_2}{\lambda_2}+\frac{\sigma_2}{2\lambda_2}}\right|\nonumber\\
&\le \frac{|G_T^\uda|\coth\frac{d_2}{\lambda_2}+\frac{\sigma_2}{2\lambda_2}}{G_2^\uda\coth\frac{d_2}{\lambda_2}+\frac{\sigma_2}{2\lambda_2}}<1.
\end{align}
Thus the numerator is also positive and $T_{\rm eff,1}>0$.

Now we calculate
\begin{equation}
1-\frac{T_{\rm eff,1}}{T_1}=\frac{\tilde{G}_T^{\uda2}\left(\frac{T_2}{T_2'}-T_{12}\right)}{1-\tilde{G}_T^{\uda2}T_{12}}.
\end{equation}
Since $T_{\rm eff,1}$ is positive, $|T_{\rm eff,1}|<T_1$ if and only if $1-T_{\rm eff,1}/T_1>0$. Since the numerator is positive, the sign of $1-T_{\rm eff,1}/T_1$ is determined by that of $T_2/T_2'-T_{12}$. Hence, we conclude that
\begin{equation}
\left.\begin{array}{c}
T_{\rm eff,1}>T_1\\T_{\rm eff,1}=T_1\\T_{\rm eff,1}<T_1
\end{array}\right\}~\text{if and only if}
\left\{\begin{array}{c}
T_{12}>T_2/T_2'\\T_{12}=T_2/T_2'\\T_{12}<T_2/T_2'
\end{array}\right.,
\end{equation}
under our assumptions.

\section{Proof of $T_{\rm 12}<T_2/T_2'$ for a positive real $G_T^\uda$\label{Sec(A):condition2}}
First we define
\begin{equation}\label{Eq(A):tildeTi}
\tilde{T}_i\equiv T_i\frac{\coth\frac{d_i}{\lambda_i}}{\tanh\frac{d_i}{2\lambda_i}}=\frac{G_i^\uda \coth\frac{d_i}{\lambda_i}}{G_i^\uda \coth\frac{d_i}{\lambda_i}+\frac{\sigma_i}{2\lambda_i}}
\end{equation}
then $T_{12}=\tilde{T}_1\tilde{T}_2$. Since $\sigma_i$ and $\lambda_i$ are positive, Eq.~(\ref{Eq(A):tildeTi}) implies that $\tilde{T}_i<1$ if $G_i^\uda$ is positive and real. Therefore, we obtain $T_{12}<1$.

Next we consider
\begin{equation}
\frac{T_2}{T_2'}=\frac{\coth\frac{d_2}{\lambda_2}+G_T^{\uda-1}\frac{\sigma_2}{2\lambda_2}}{\coth\frac{d_2}{\lambda_2}+G_2^{\uda-1}\frac{\sigma_2}{2\lambda_2}}>1,
\end{equation}
if $G_T^\uda$ is positive and smaller thatn $G_2^\uda$. As a result, we obtain $T_{\rm 12}<1<T_2/T_2'$

\end{appendix}


\end{document}